\def \SAIT #1 #2 {{\em Mem.\ Soc.\ Astron.\ It.\/} {\bf #1}, #2}
\def \MESS #1 #2 {{\em The Messenger\/} {\bf #1}, #2}
\def \ASTRNACH #1 #2 {{\em Astron. Nach.\/} {\bf #1}, #2}
\def \AAP #1 #2 {{\em Astron. Astrophys.\/} {\bf #1}, #2}
\def \AAL #1 #2 {{\em Astron. Astrophys. Lett.\/} {\bf #1}, L#2}
\def \AAR #1 #2 {{\em Astron. Astrophys. Rev.\/} {\bf #1}, #2}
\def \AAS #1 #2 {{\em Astron. Astrophys. Suppl. Ser.\/} {\bf #1}, #2}
\def \AJ #1 #2 {{\em Astron. J.\/} {\bf #1}, #2}
\def \ANNREV #1 #2 {{\em Ann. Rev. Astron. Astrophys.\/} {\bf #1}, #2}
\def \APJ #1 #2 {{\em Astrophys. J.\/} {\bf #1}, #2}
\def \APJL #1 #2 {{\em Astrophys.. J. Lett.\/} {\bf #1}, L#2}
\def \APJS #1 #2 {{\em Astrophys. J. Suppl.\/} {\bf #1}, #2}
\def \APSS #1 #2 {{\em Astrophys. Space Sci.\/} {\bf #1}, #2}
\def \ASR #1 #2 {{\em Adv. Space Res.\/} {\bf #1}, #2}
\def \BAIC #1 #2 {{\em Bull. Astron. Inst. Czechosl.\/} {\bf #1}, #2}
\def \JSQRT #1 #2 {{\em J. Quant. Spectrosc. Radiat. Transfer\/} {\bf #1}, #2}
\def \MN #1 #2 {{\em Mon. Not. R. Astr. Soc.\/} {\bf #1}, #2}
\def \MEM #1 #2 {{\em Mem. R. Astr. Soc.\/} {\bf #1}, #2}
\def \PLR #1 #2 {{\em Phys. Lett. Rev.\/} {\bf #1}, #2}
\def \PASJ #1 #2 {{\em Publ. Astron. Soc. Japan\/} {\bf #1}, #2}
\def \PASP #1 #2 {{\em Publ. Astr. Soc. Pacific\/} {\bf #1}, #2}
\def \NAT #1 #2 {{\em Nature\/} {\bf #1}, #2}

\documentstyle{memsait}
\input epsf.sty

\def\lesssim{\mathrel{\hbox{\rlap{\hbox{\lower4pt\hbox{$\sim$}}}\hbox{$<$}}}}
\def\gtrsim{\mathrel{\hbox{\rlap{\hbox{\lower4pt\hbox{$\sim$}}}\hbox{$>$}}}}

\begin{opening}
\title{ GRBs and the Ghost of the Fireball }
\author{Daniele Fargion$^1$}
\institute{$^1$Physics Department and INFN,Rome,Italy\\
}

\date{} 
\end{opening}
\begin{document}
%
%
%

\oddpagefooter{}{}{} 
\evenpagefooter{}{}{} 
\
\bigskip

%
%


\begin{abstract}
 Gamma Ray Burst has been widely believed in last decade to
be super-explosions: the Fireball. We are argue on the contrary
that GRBs (as well as Soft Gamma Repeaters SGR) are precessing
Gamma Jets. We remind the list of contradiction that Fireball and
its galactic smaller version, the magnetar,have to face. In
particular the existence of weak isolated X-ray precursor signal
before the main Gamma Ray Burst and (rare SGR) events disagree
with any explosive, one shoot, scenarios either isotropic or
wide-beamed. We interpret them as earlier marginal blazing of
outlying X conical Jet tails of precessing, spinning $\gamma$ Jet.
\end{abstract}

\section{Introduction: GRBs and the blow up of the energy}
Gamma Ray Burst, GRB, has been associated in last decade with huge
explosions many order of magnitude ($10^{10}$) more power-full
than common Supernova. These bursts are so intense ( because of
GRB cosmic distance origin) and energetic ( at the electron mass
edges) as well as so sharp (within millisecond scale in time) that
their own opacity (at corresponding small hundred kilometers
sizes) make difficult their own shining outside (an over
Eddington luminosity). This Eddington-opacity is reached also for
smaller distance (and less power-full) sources, named Soft Gamma
Repeaters SGRs, because softer and sometime repeating. For all
last decade ruled Fireball: a huge isotropic explosive model where
boosted relativistic electron pairs sea and their annihilation may
have remitted soft KeV photons into harder MeV energy band in a
wide shells. The GRBs structure arises in Fireball by
internal-external shell scattering. We considered since 1993 the
$\gamma$ Jet model for GRBs and SGRs both of them at largest
galactic halo; after the rare GRB 980425 and its Supernova-GRB
connection, we committed to explain both of them as a precessing
Jet at different energy regime, both for GRBs at Supernova beamed
power and  SGRs at much lower X-pulsar beamed luminosity. Most of
the scientific community remained firmly on Fireball model before
and  after the first cosmic identification of GRB at cosmic edges
on February 1997. For me in that epoch was difficult to accept
huge supernova jet and I preferred a galactic halo model; since
April 25 GRB, however, I immediately believed in GRB and SGR
unified model at huge different beamed powers. However the
fireball believers, a formidable school all over the world,
increased their  excitement as much as the GRB distance (or
redshift) and its apparent output power  was increased by event
to event. Their explosive model was feed by the apparent X-decay
after-glow tail.  Nevertheless the huge $GRB990123$ event blow up
too much even for the most excited mind: its unbelievable energy
puzzle (imagine, two or more solar masses totally converted in
pure gamma  within tens of seconds within a millisecond-hundred
kilometer substructures) forced most authors to bend their
collimate Fireball model toward en-longed beamed Jet, in order to
economize the energy. Therefore with a polite compromise the
Fireball models shifted toward  a wide (ten degrees or so ) open
cone or a Fountain Jet Models. Still one-shoot Fireball model.
How was possible to hold slimy neutrino in such a Jet is a
mystery. No general agreement on the Supernova-GRB connection was
in general acceptable to Fireball believers. The new Jet-Fireball
shared still the explosive nature and the shock wave time
modulation ruled by external relic shells, but it relaxed the
energy demand by the inverse of the solid angle beam. The word
beam or jet has been often hidden under new fashionable names. As
the time went on the beam of Fireball's model shrinks and the
GRBs energy reduces from ($\gg$ $10^{54}$ $erg $) first to ($\gg$
$10^{51}$ $erg $) and even toward ($\gg$ $10^{48}$ $erg $).
However the smaller the (explosive) beam angle the higher is the
needed GRB event rate: from nominal $10^{-6}$ event a year for
galaxy one must expect a more frequent $10^{-3}$ or even $1$ GRB a
year in galaxies. Therefore we must accept the co-existence in
our Universe of two totally separate wild animals: Supernova whose
explosive isotropic output reaches $10^{44}$ $erg~s^{-1}$ powers
and a total energy $10^{51}$ $erg$ and a as common as animal
whose jet output is  $10^{48}$ $erg~s^{-1}$ and total energy
$10^{51}$ $erg$. A possible GRB-Supernova connection call for an
additional paradox: Why GRB Fireball-Jet power is 4-7 order of
magnitude higher than the corresponding optical one in Supernova.
Why not to respect at least energy equi-partition? In this
Fireball scenario the rare GRB of April 1998 stand outside or
better is often hidden or taken away because of the more puzzling
apparent $low$ energy output and softer nature while being nearer
and un-red-shifted. Of course off-axis Fireball event make the
miracle but cannot solve the statistics.\\
 We argued (1998-1999) on the contrary that GRBs and SGRs find a comprehensive theory
within a very thin (tens of seconds) spinning
  and multi precessing $\gamma$ Jet, sprayed by a Neutron Star, NS, or Black Hole, BH,
  (Fargion 1994-1999, Fargion, Salis 1995-1998).
   Higher black holes mass ( $10^{1}$ $10^{4}$ solar masses)
cannot help  Fireball isotropic survival because contradictions
with the sharp GRB time-scales.
    We have shown that the same rarity of GRB-SN detection
  and the established GRB980425-SN link favors very clearly
   the thin (tens of arc second or millisecond radiant) Jet Nature of GRBs. Such thin jet
   may accommodate the Supernova power into an apparent GRB
   luminosity in agreement with energy equi-partition.
    Moreover the Jet in precession explains Soft Gamma Repeaters.
    Even originally ($1970-80$) most authors unified  GRB/SGR models ,
   since last fifteen years  are commonly separated by their repeater and spectra differences
   and finally their much nearer galactic distances;
 however we and other have shown that they, rarely, openly shared
  the same spectra, time and flux structures.
   Last SGR1900+14 (May-August-October 1998) events and SGR1627-41
(June-October 1998) events did exhibit at peak intensities hard
spectra comparable with classical GRBs. Indeed  the SGR1900+14
event BATSE   trigger 7171 left an almost  identical event
comparable to a just following  GRB (trigger 7172) on   the same
day, same detector, with same spectra and comparable flux (Fargion
1998-1999) . This Hard-Soft connection has been re-discovered
and confirmed more recently  by BATSE group ( Woods et all  1999)
 with an additional hard event of SGR 1900+14 recorded in GRB990110 event.\\
    Additional GRB-SGR connection occur between  GRB980706 event with an almost identical (in time, channel spectra, morphology and
   intensities) observed in GRB980618 originated by SGR 1627-41. Nature would be   very perverse in mimic two signals,
   (even  if at different distances and different powers),  by two extreme different source engines.
  This points once again to their common Nature.
  At least SGRs offer the unique occasion for a study of a near-by laboratory
  GRB to a better understanding of both models. However they are both
  up present times usually described by catastrophic spherical explosions,
  Fireball and Magnetar Models.
   Their different distances, cosmic versus galactic ones, imply
very different power source Jet, but their morphological
similarity strongly suggest an unique process: the blazing of  a
spinning and multi-precessing gamma Jet, from either Neutron Star
or Black Hole. The $\gamma$ Jet is born by high GeVs electron
pairs Jet which are regenerating, via Inverse Compton Scattering,
an inner collimated beamed $\gamma$ (MeVs) precessing  Jet. The
thin jet (an opening  angle inverse of the electron Lorentz
factor, a milli-radiant or below), while spinning, is driven by a
companion and/or an asymmetric accreting disk in a Quasi Periodic
Oscillation (QPO) and in a Keplerian multi-precessing blazing
mode: its $\gamma-X$ ray lighthouse trembling and flashing is the
source of the complex and wide structure of observed Gamma Bursts.
These $\gamma$ Jets share a peak power of a Supernova ($10^{44}
erg s^{-1}$) at their birth (during SN and Neutron Star
formations), decaying by power law $\sim t^{-1}$ $-$ $\sim
t^{-(1.5)}$ to less power-full Jets that converge to present
persistent SGRs stages. Indeed these ones  are blazing events
from late relic X pulsar observable only at nearer distances. The
$\gamma$ Jet emit in general at $\sim$ $ 10^{35}$ erg$ s^{-1}$
powers; both of GRB and SGR show an apparent luminosity amplified
by the inverse square of the thin from $10^{-3}$ to $10^{-4}$
radiant angle Jet beaming: the corresponding solid angle $\Omega$
spreads between $10^{-7}$ and $10^{-9}$ leading to the necessary
amplification for both SN-GRB and X pulsar-SGR . Optical-Radio
After-Glows are not the fading fireball explosion tails often
observed in puzzling variable non monotonic decay, but the
averaged external Jet tails moving and precessing and
geometrically fading away. The rare optical re-brightening (the
so called SN bump) observed in few afterglow has been, very
probably, erroneously associated to an underlying isotropic SN
flash: the optical re-brightening  afterglow is in general the
late re-crossing of the precessing Jet tail periphery toward the
observer direction. The averaged integral optical signal (within
thousand of seconds) hide the short-scale oscillatory behaviour
of the precessing Jet. In particular the geometrical beaming
offered in the rare GRB970508 a peculiar optical re-brightening
and a manifest late radio oscillating afterglow of the cycloidal
lighthouse   Jet.
\section{The  geometrical multi-precessing Gamma Jet and GRB bursting signature}
 We imagine the GRB and SGR
nature as the early and the late stages of jets fueled first by
SN event and later by a disk or a companion (WD, NS) star. Their
binary angular velocity $\omega_b$ reflects the beam evolution
$\theta_1(t) = \sqrt{\theta_{1 m}^2 + (\omega_b t)^2}$ or more
generally a multi-precessing angle $\theta_1(t)$ (Fargion \&
Salis 1996):
\begin{equation}\label{eq7}
  \theta_1(t) = \sqrt{\theta_{x}^2 +\theta_{y}^2 }
\end{equation}
\begin{equation}\label{eq8}
  \theta_{x}(t) =                               
  \theta_{b} sin(\omega_{b} t + \varphi_{b})+
  \theta_{psr}sin(\omega_{psr} t)+
  \theta_{N}sin(\omega_{N} t  + \varphi_{N})
\end{equation}

\begin{equation}\label{eq8}
  \theta_{y}(t) = \theta_{1 m}+
  \theta_{b} cos(\omega_{b} t + \varphi_{b})+
  \theta_{psr} cos(\omega_{psr} t)+
  \theta_{N} cos(\omega_{N} t  + \varphi_{N})
\end{equation}
where $\theta_{1 m}$ is the minimal angle impact parameter of the
jet toward the observer, $\theta_{b}$, $\theta_{psr}$,
$\theta_{N}$ are, in the order, the maximal opening precessing
angles due to the binary, spinning pulsar, nutation mode of the
jet axis. For a 3D pattern and its projection along the vertical
axis in an orthogonal 2D plane (Fig.1)
The angular velocities combined in the multi-precession keep
memory of the pulsar jet spin ($\omega_{psr}$), the precession by
the binary $\omega_b$ and an additional nutation due to inertial
momentum anisotropies or beam-accretion disk torques
($\omega_N$). On average, from eq.(5) the $\gamma$ flux and the
$X$ optical afterglow decays, in first approximation, as $t^{-2}$;
the complicated spinning and precessing jet blazing is responsible
for the wide morphology of GRBs and SGRs as well as their
internal periodicity.(See Fig.2). The consequent $\gamma$ time
evolution and spectra derived in this ideal models may be
compared successfully with similar observed GRB data evolution as
shown in Fig. 3, regarding X-ray precursors in $GRB 971210,
GRB971212, GRB990518, GRB000131$.
\begin{figure}
\epsfysize=7cm
\hspace{2.0cm}\epsfbox{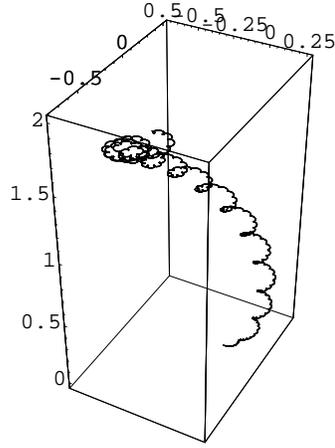} 
\caption[h]{Leth:Three dimensional space evolution of the
Precessing  Gamma Jet leading, by its blazing
  to X precursor and main GRBs. Right: the same pattern observed from above
  along the vertical axis , in the orthogonal two dimensional plane}
  \end{figure}
\begin{figure}
 \epsfysize=7cm
\hspace{2.0cm}\epsfbox{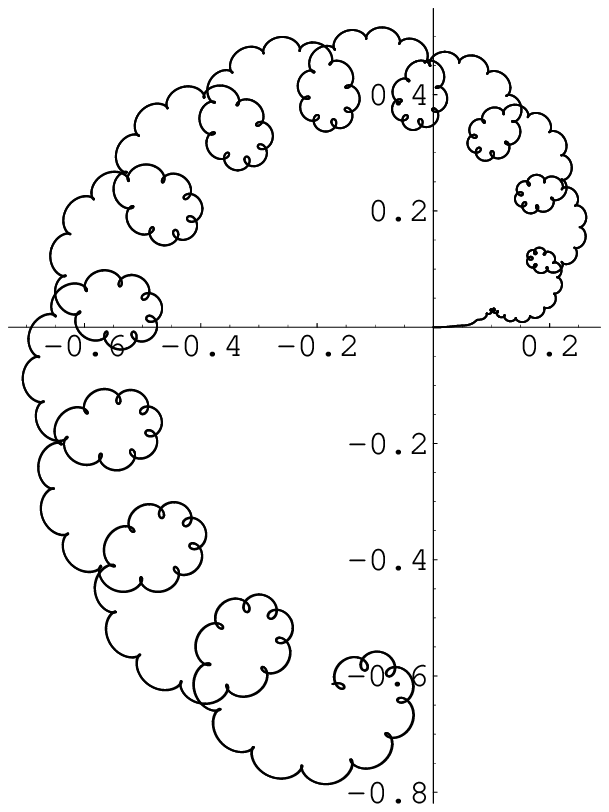} 
\caption[h]
 {Leth:Three dimensional space evolution of the Precessing
  Gamma Jet leading, by its blazing
  to X precursor and main GRBs. Right: the same pattern observed from above
  along the vertical axis , in the orthogonal two dimensional plane}
 \label{eps4}
\end{figure}
\begin{figure}[h]
\epsfysize=7cm
\hspace{2.0cm}\epsfbox{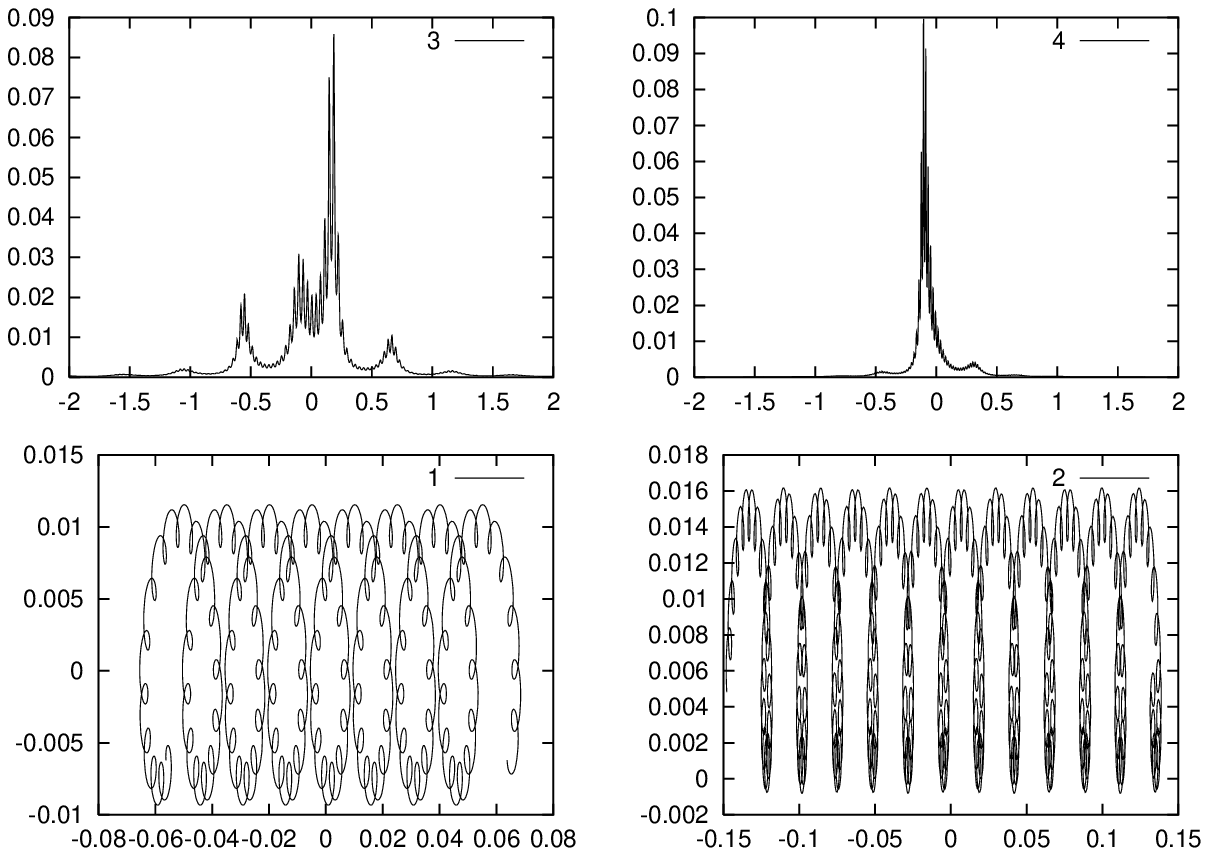} 
\caption[h]{Down : Label 1 and 2, in two different
  bi-dimensional angle Spinning, Precessing Gamma Jet ring patterns toward the
   detector at the  origin ($0,0$) corresponding to previous like 3D precessing jet.
   Up: Label 3-4, the consequent X, $\gamma$ intensity time evolution
   signals derived by ICS formula. X ray precursor may naturally arise in some configurations.}
  \label{eps5}
\end{figure}
 Similar descriptions with more parameters and with a rapid time evolution
of the jet has been more recently also proposed by (Portegies
Zwart et all 1999).
\section{ X Ray precursor as the fingerprint of Precessing Gamma Jets and the end of Fireballs}
   To choose for a  model  let us just consider with no prejudice
   the  last reported (and most distant $z= 4.5$) event:
   GRB000131 and its $X$ ray precursor:(fig 3,last).
   This event while being red-shifted and slowed down by a
   factor 5.5 exhibit on the contrary a short scale time fine structure not explicable
   by any fireball model  , but  well compatible  with
    a thin, fast spinning precessing $\gamma$ jet.
   The extreme $\gamma$ energy budget, calling for a
   comparable $\nu$ one, exceeds few solar masses in its main
   emission even for ideal full energy conversion.
    Moreover one must notice the presence of a weak $X$-ray precursor pulse
   lasting 7 sec, 62 sec before the huge main structured $\gamma$ burst
   trigger. Its arrival direction (within 12 degree error) with main GRB
   is consistent only with the main pulse (a probability to occur by chance below $3.6
   10^{-3}$).   The time clustering proximity (one minute over a day GRB rate average) has the probability
   to occur by chance below once over a thousand.   The over all probability to observe this precursor by change is below 3.4 over a million
   making inseparable its association with the main  GRB000131 event.
    This weak burst signal correspond to a power above a million
   Supernova and have left no trace or Optical/X transient just a minute
   before the real (peak power $> billion $ Supernova) energetic event.
   Similar X precursors occurred in a non negligible minor sample of GRBs (see for example Fig
   2-4a) and also few SGRs event (Fig 4b).   No isotropic GRB explosive progenitor could survive such a disruptive
   isotropic(million supernova output)   precursor trigger nor any multi-exploding jet. Only a persistent precessing Gamma Jet
    crossing nearby the observer direction twice could naturally explain it.
\begin{figure}[h]
 \epsfysize=7cm \hspace{2 cm}\epsfbox{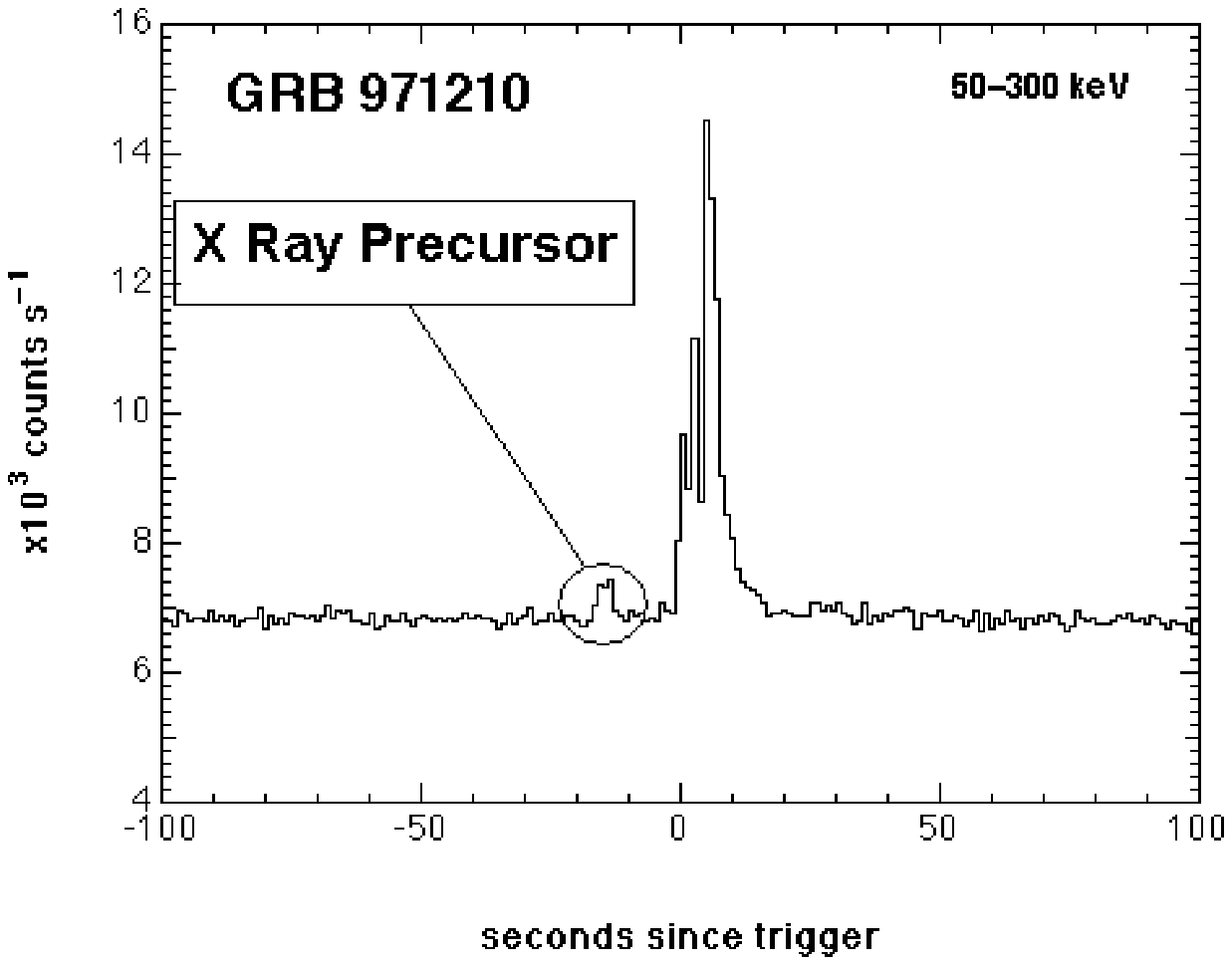}
\caption[]{Up: Time evolution and X precursors in GRB $971210$
\label{eps1}}
\end{figure}
\begin{figure}[h]
\epsfysize=7cm \hspace{2 cm}\epsfbox{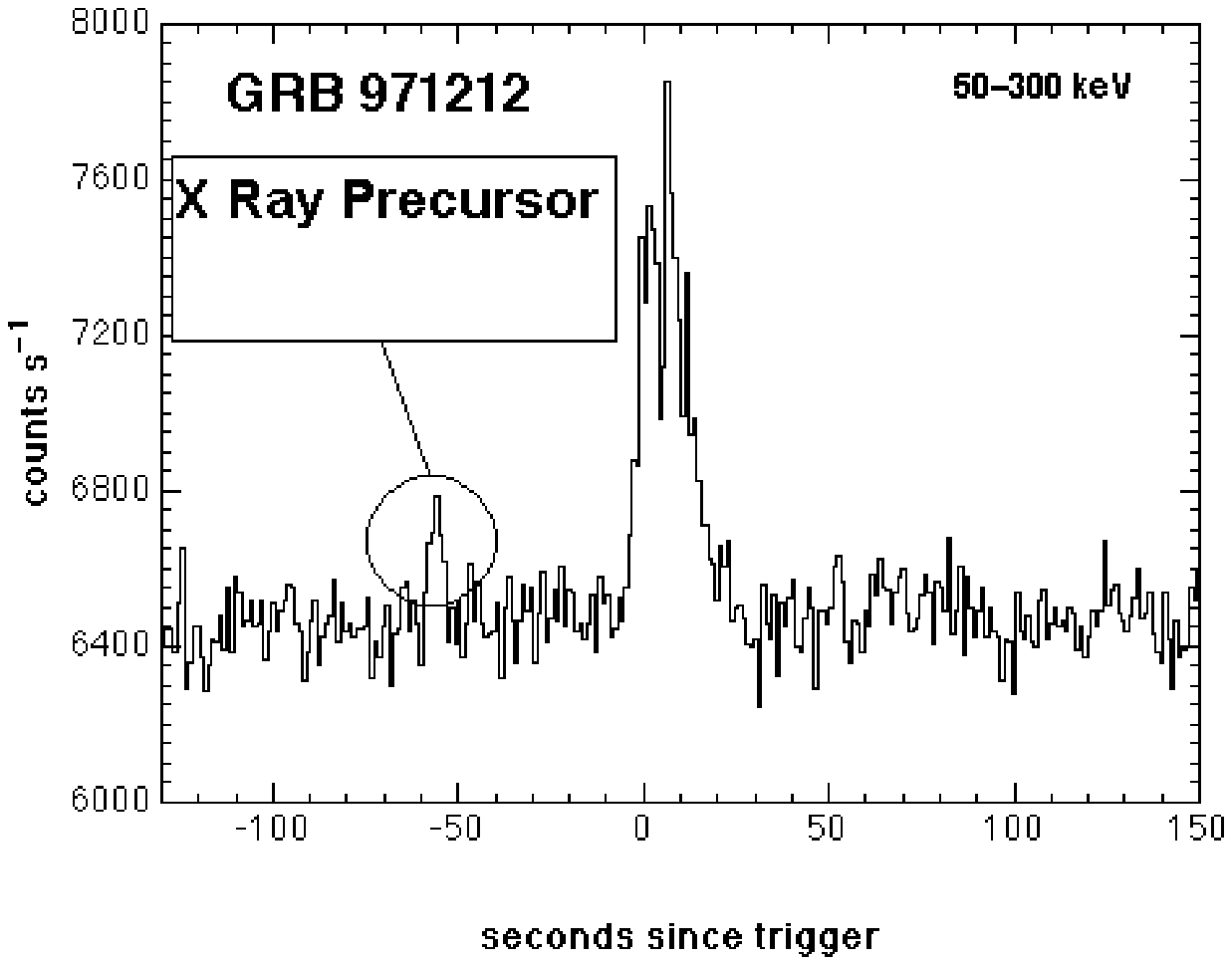} \caption[]{Up:
Time evolution and X precursors in GRB $971212$.  \label{eps1}}
\end{figure}
\begin{figure}[h]
\epsfysize=7cm
\hspace{2.0cm}\epsfbox{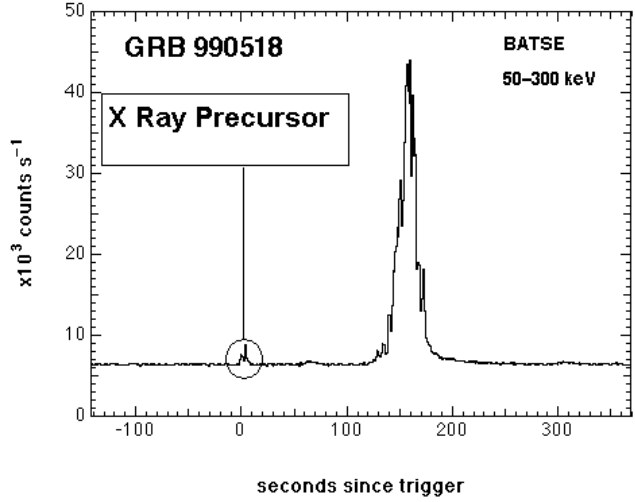}
 \caption[]{ Time evolution and X precursors in GRB $990518$.  \label{eps1}}
\end{figure}
\begin{figure}[h]
\epsfysize=7cm \hspace{2.0cm}\epsfbox{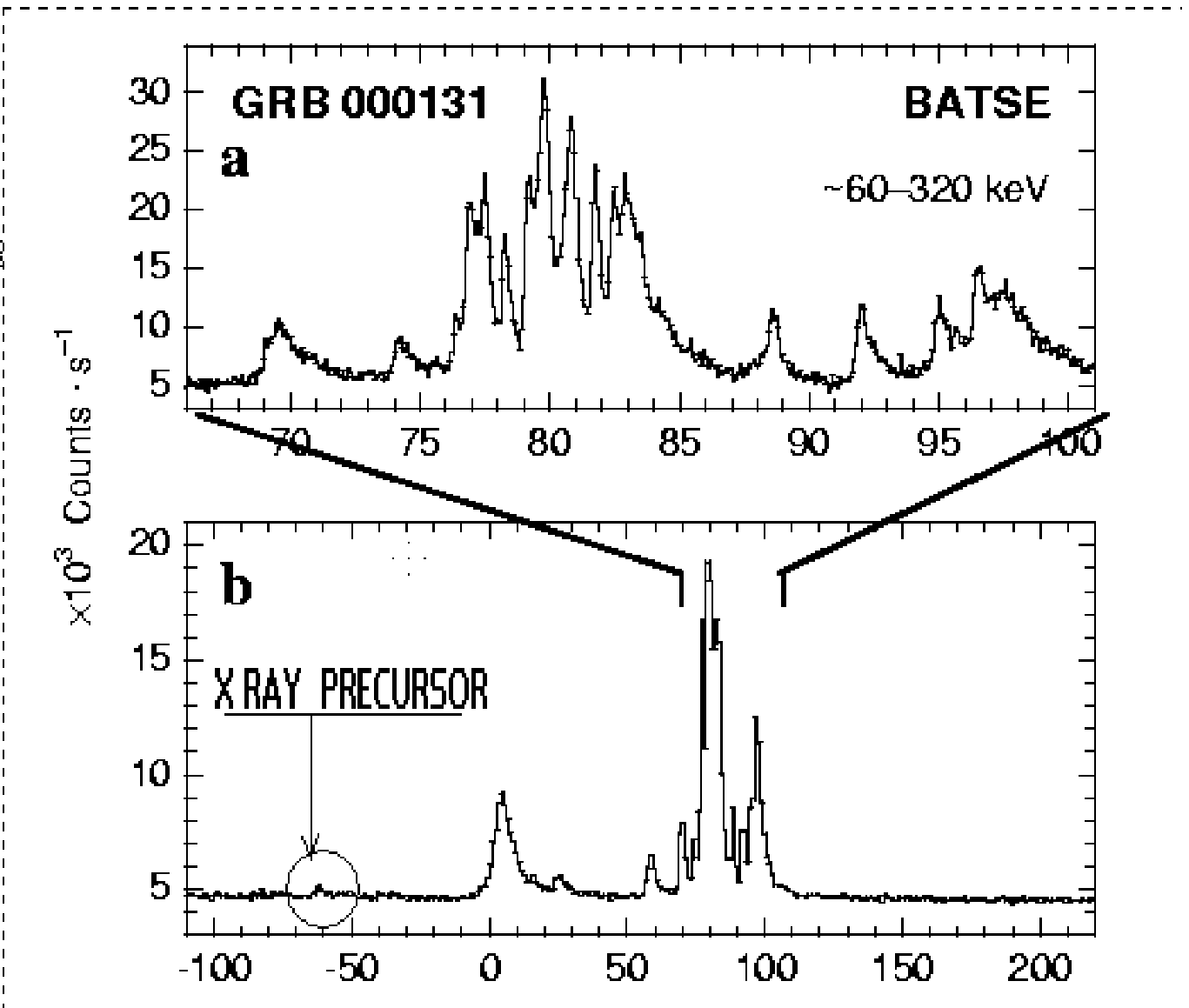} \caption[]{ Time
evolution and X precursors in most distant (red-shift 4.5) GRB
$000131$.  \label{eps1}}
\end{figure}




\section{  Thin Precessing $\gamma$ Jet and the GRB-SN Connection  }

Statistical arguments (Fargion 1998, 1999) favor a unified GRBs
model based on blazing, spinning and precessing thin jet. We
assume that GRB jet arise in most SNe outbursts. The far GRBs are
observables at their peak intensities (coincident to SN) while
blazing in axis to us within the thin jet very rarely;
consequently the hit of the target occurs only within a wide
sample of sources found in  a huge cosmic volume. In this frame
work the GRB rate do not differ much from the SN rate. Assuming a
SN-GRB event every 30 years in a galaxy and assuming a thin
 angular cone ($\Omega < $(1/4)$10^{-8}$) the probability to be within
 the cone jet in a ($ 4*10^{10}$)cosmic  sample of galaxies
  (at limited Hubble R$\geq$ 28 magnitude ) within
 our main present observable Universe volume ($z \sim 1$-$z \sim 4$)
 during one day of observation at a nominal 10 sec GRB
 duration is quite small: ($P < 10^{-2}$). This value should be suppressed
 by nearly an order of magnitude because of the detector acceptance. However a precessing
 gamma jet whose decaying scale time is a thousand time longer than the GRB itself
 (decaying by a power law $\sim t^{-1}$)
 whose scale time is  nearly  ten or twenty thousand of seconds,
  may fit naturally the observed GRB rate. Moreover the jet pressure could also accumulate gas and form dense
filaments. Such filaments fragment as well as  gravitationally
cluster leading to contemporaneous  stellar arc
formations.(Y.Efremov\&D.Fargion, 2000). This phenomena might
explain why GRBs seem to be associated with star formation
regions.

\section{GRBs  Energy Budget in Fireball and the Neutrino $\nu$ and $\gamma$ ejection}

Gamma Ray Bursts as recent $GRB990123$ and $GRB990510$ emit, for
isotropic explosions, $\gamma$ energies as large as  two solar
masses annihilation. These energies are underestimated because of
the neglected role of comparable ejected MeV (Comptel signal)
neutrinos $\nu$ bursts and assume an unrealistic ideal energy
conversion efficiency. Indeed, as often neglected, it is
important to remind that the huge energy bath (for a fireball
model) on GRB990123 imply also a corresponding neutrino burst. As
in hot universe, if entropy conservation holds, the $\nu$ energy
density factor to be added to the photon $\gamma$  budget is at
least $( \simeq (21/8)\times (4 /11)^{4/3} )$. If entropy
conservation do not hold the energy needed is at least a factor
$[21/8]$ larger than the gamma one. The consequent total
energy-mass needed for the two cases are respectively 3.5 and 7.2
solar masses. Additional factors must be introduced for realistic
energy conversion efficiency leading to energies as large as tens
of solar masses. No fireball by NS may coexist with it. Jet could.
 These extreme power cannot be explained with any
standard spherically symmetric Black Hole Fireball model. A too
heavy Black Hole (hundred or thousands solar masses) or, worse,
Star would be unable to coexist with the shortest millisecond time
structure of Gamma Ray Burst. Cosmological and nearby
Gravitational Red-shifts may only make the Fireball Model more
inconsistent. Smaller size BH or NS do not offer enough mass
reserve. Beaming of the gamma radiation may overcome the energy
puzzle along with the short scale-time. However any mild
''explosive beam'' event as some models (Wang \& Wheeler 1998)
$(\Omega > 10^{-2} )$ would not solve the jet containment at the
corresponding disruptive energies. Moreover such a small beaming
would not solve the huge GRBs flux energy windows ($10^{47} \div
10^{54}$ erg/sec), keeping GRB980425 and GRB990123 within the
same GRB framework.
 Only extreme beaming $(\Omega \sim 10^{-8} )$, by a
slow decaying, but long-lived precessing jet, may coexist with
characteristic Supernova energies, apparent GRBs output and the
puzzling GRB980425 statistics as well as the GRB connection with
older, nearer and weaker SGRs relics. Therefore SGRs are very
useful nearby astrophysical Laboratory where to study and test
the far GRB process. SGRs are not associated with huge OT
afterglow or explosive SN event : they have lost their primordial
SNR shells elsewhere while escaping at high velocity far from the
SN birth place.
\section{Hard Gamma Jet by Inverse Compton Scattering by GeV Electron Pairs beam}

 A jet angle related by a
relativistic kinematics would imply $\theta \sim
\frac{1}{\gamma_e}$, where $\gamma_e$ is found to reach $\gamma_e
\simeq 10^3 \div 10^4$ (Fargion 1994, 1998). At first
approximation the gamma constrains is given by Inverse Compton
relation: $< \epsilon_\gamma > \simeq \gamma_e^2 \, k T$ for $kT
\simeq 10^{-3}-10^{-1}\, eV$ and $E_e \sim GeVs$ leading to
characteristic X-$\gamma$ GRB spectra.  The origin of $GeVs$
electron pairs are  very probably decayed secondary related to
primary inner muon pairs jets, able to cross dense stellar target.
The consequent adimensional photon number rate (Fargion \& Salis
1996) as a function of the observational angle $\theta_1$
responsible for peak luminosity  becomes
\begin{equation}
\frac{\left( \frac{dN_{1}}{dt_{1}\, d\theta _{1}}\right) _{\theta
_{1}(t)}}{ \left( \frac{dN_{1}}{dt_{1}\, d\theta _{1}}\right)
_{\theta _{1}=0}}\simeq \frac{1+\gamma ^{4}\, \theta
_{1}^{4}(t)}{[1+\gamma ^{2}\, \theta _{1}^{2}(t)]^{4}}\, \theta
_{1}\approx \frac{1}{(\theta _{1})^{3}} \;\;.\label{eq4}
\end{equation}
The total fluence at minimal impact angle $\theta_{1 m}$
responsible for the average luminosity  is
\begin{equation}
\frac{dN_{1}}{dt_{1}}(\theta _{1m})\simeq \int_{\theta
_{1m}}^{\infty }\frac{ 1+\gamma ^{4}\, \theta _{1}^{4}}{[1+\gamma
^{2}\, \theta _{1}^{2}]^{4}} \, \theta _{1}\, d\theta _{1}\simeq
\frac{1}{(\, \theta _{1m})^{2}}\;\;\;. \label{eq5}
\end{equation}
These spectra fit GRBs observed ones (Fargion \& Salis 1995).
Assuming a beam jet intensity $I_1$ comparable with maximal SN
luminosity, $I_1 \simeq 10^{45}\;erg\, s^{-1}$, and replacing
this value in the above a-dimensional   equation  we find a
maximal apparent GRB power for beaming angles $10^{-3} \div
3\times 10^{-5}$, $P \simeq 4 \pi I_1 \theta^{-2} \simeq 10^{52}
\div 10^{55} erg \, s^{-1}$, just within observed ones. We also
assumed a power law jet time decay as follows
\begin{equation}\label{eq6}
  I_{jet} = I_1 \left(\frac{t}{t_0} \right)^{-\alpha} \simeq
  10^{45} \left(\frac{t}{3 \cdot 10^4 s} \right)^{-1} \; erg \,
  s^{-1}
\end{equation}
where ($\alpha \simeq 1$) is a value able to reach, at 1000 years
time scales, the present known galactic micro-jet (as SS433)
intensities powers: $I_{jet} \simeq 10^{39}\;erg\, s^{-1}$. This
offer a natural link between the GRB and the SGR out-put powers.
We used the model to evaluate if April precessing jet might hit us
once again. It should be noted that a steady angular velocity
would imply an intensity variability ($I \sim \theta^{-2} \sim
t^{-2}$) corresponding to some of the earliest afterglow decay
law.

\section{Precessing Jet  Relics in action}
  The Gamma Jet progenitor of the GRB
   is leaving a trace in the space: usually a nebulae
where  the nearby  ISM  may record the jet  sweeping as a
     three dimensional screen. The outcomes  maybe either
      a twin ring as recent SN1987A has shown, or helix traces as
       the Cat Eye Nebula or more structured shapes as plerions
        and hourglass nebulae.
         We imagine the jet as born by a binary system (or
         by an asymmetric disk accreting interaction)
         where the compact companion (BH or NS) is the source of the
         ultra relativistic electron pair jet (at tens GeV.
       Inverse Compton Scattering on IR thermal photons will produce a collinear
        gamma jet at MeV). The rarest case where the jet
         is spinning and nearly isolated would produce a jet train
         whose trace are star chains as the Herbig Haro ones
         (Fargion, Salis 1995). When the jet
         is modified  by the magnetic field torque of the
         binary companion field the result may be a more rich cone shape.
         If the ecliptic  lay on the same plane orthogonal to the jet
         in an ideal circular orbit than the bending
         will produce an ideal twin precessing cones which is
           reflected in an ideal twin rings (Fargion, Salis 1995).
         If the companion is in eccentric orbit the resultant
         conical jet will be more deflected at perihelion while
          remain nearly undeflected at a aphelion.
          The consequent off-axis cones will play the role of a
          mild "rowing" acceleration  able to move the system
          and speed it far from its original birth (explosive)
          place. Possible traces are the asymmetric external twin
          rings painted onto the spherical relic shell by SN1987a.
          Fast relics NS may be speeded by this processes (Fargion, Salis
          1995a, 1995b, 1995c). Because of momentum conservation this asymmetric
            rowing is the source of a motion of the jet relic  in the
            South-East direction. In extreme eccentric system
            the internal region of the ring are more  powered
            by the nearby encounter leading to the apparent gas arcs.
            If the system is orbiting in a plane different from
              the one orthogonal to the jet the outcoming
              precessing jet may spread into a mobile twin cone
              whose filling may appear as a full cone or a twin hourglass
              by   a common plerion shape. At late times there is also  possible
               apparent spherical shapes sprayed and structured by a chaotic
               helix.  External  ISM distribution may also play a role enhancing
                 some sides or regions of the arcs.
                 The same presence of external relic shells may
                 explain the possible new evidences of iron lines
                 observed by Beppo-Sax satellite. We may explain
                 these rare signal as the re-illuminated isotropic
                 emission due to the under-line beamed inner
                 light-house $\gamma$   Jet.
              The integral jet spray in  long times may mimic even
                spherical envelopes but internal detailed inspection
                might soon  reveal the thinner jet origin (as in recent Eta
                Carina string jets). Variable nebulae behaviours recently
                observed are confirming our present scenario.

\section{Conclusions}
Gamma Ray Burst call for extreme isotropic power or extreme
collimation or both. Fore a decade the isotropic Fireball story
hold the market. Now a Fireball-Jet compromise start to  float on
the community. Like a ghost. \\
 We strongly believe that GRBs and SGRs  are persistent blazing flashes from light-house
thin $\gamma$ Jet spinning  in multi-precessing (binary,
precession, nutation)  mode. These Jets are originated by NSs or
BH in binary system or disk powered by infall matter: the Jet is
not a single explosive event even in GRB, but they are powered at
maximal output during SN event.The Jet power is comparable to SN
at its peak;  the $\gamma$ Jet has a chain of progenitor
identities: it is born in most SN and or BH birth and it is very
probably originated by very collimated primary muon pairs at
GeVs-TeVs energies. These muons could cross the dense target
matter around the SN explosions, nearly transparent to
photon-photon opacities. These muon progenitors might be
themselves secondary relics of pion decays or even by a more
transparent beamed ultra-high energy neutrino Jet originated (by
hadronic and pion showering) near the NS or BH. The high energy
relativistic muons  decay in flight in electron pairs is itself
source of GeVs relativistic pairs whose Inverse Compton
Scattering with nearby thermal photon is the final source of the
observed  hard $X$ - $\gamma$ Jet. The relativistic morphology of
the Jet and its multi-precession is the source of the puzzling
complex $X$-$\gamma$ spectra signature of GRBs and SGRs. Its
inner internal Jet contain, following the relativistic Inverse
Compton Scattering,  hardest and rarest beamed GeVs-MeVs photons
(as the rarest and long life  EGRET GRB940217 one) but its
external Jet cones are dressed by softer and softer photons.
This   onion like multi Jets is not totally axis symmetric: it
doesn't appear in inner structure, on its front, as a concentric
ring serial; while turning and spraying around it is deformed
(often) into an elliptical off-axis concentric rings preceded by
the internal harder center leading to a common hard to soft GRBs
(and SGRs) train signal. In our present model and simulation this
internal effect has been here neglected without any major
consequence. The complex variability of GRBs and SGRs are
simulated successfully by the equations and the consequent
geometrical beamed Jet blazing leading also to the observed rare
$X-\gamma$ signatures. As shown in fig 2 the slightly different
precessing configurations could easily mimic the wide morphology
of GRBs as well as the surprising rare X-ray precursor shown in
Fig. 3  above.

%

\end{document}